\documentclass[
reprint,
 amsmath,amssymb,
 aps,
prd,
tightenlines
]{revtex4-1}
\usepackage{siunitx}
\usepackage{booktabs}
\usepackage{color}
\usepackage{multirow}
\usepackage{overpic}
\usepackage{colortbl}
\usepackage{graphicx}
\usepackage{float}
\usepackage{array}
\usepackage{makecell}
\usepackage{subfigure}
\usepackage{graphicx}
\usepackage{dcolumn}
\usepackage{bm}
\usepackage{xcolor}
\usepackage{amsmath}
\usepackage[T1]{fontenc}
\definecolor{aa}{RGB}{0,0,139}
\newcommand{\etap}{\eta^{\prime}}
\newcommand{\psipptophietap}{$\psi(3770)\to\phi\etap$}
\newcommand{\psiptophietap}{$\psi(2S)\to\phi\etap$}

\newcommand{\eetophietap}{$e^{+}e^{-}\to\phi\etap$}

\newcommand{\eetorhoetap}{$e^{+}e^{-}\to\rho\etap$}

\newcommand{\KpKm}{K^+K^-}
\newcommand{\gampipi}{\gamma\pi^{+}\pi^{-}}
\newcommand{\etapipi}{\eta\pi^{+}\pi^{-}}
\newcommand{\etaptogammapipi}{$\etap\to\gamma\pi^{+}\pi^{-}$}
\newcommand{\etaptoetapipi}{$\etap\to\eta\pi^{+}\pi^{-}$}
\newcommand{\etatogammagamma}{$\eta\to\gamma\gamma$}
\newcommand{\phitokk}{$\phi\to K^{+}K^{-}$}

\newcommand{\EE}{e^+e^-}
\newcommand{\TT}{$\tau^{+}\tau^{-}$}
\newcommand{\GG}{$\gamma\gamma$}

\newcommand{\eff}{\varepsilon}
\newcommand{\BR}{{\cal B}}

\newcommand{\pspp}{\psi(3770)}

\newcommand{\MM}{$\mu^+\mu^-$}

\usepackage{relsize}
\def\babar{\mbox{\slshape B\kern-0.1em{\smaller A}\kern-0.1em
    B\kern-0.1em{\smaller A\kern-0.2em R}}}

\usepackage{multirow}
\usepackage{enumerate}
\usepackage{amsmath}
\begin{document}


\title{\boldmath Measurement of \eetophietap~cross sections at center-of-mass energies from 3.508 to 4.951~GeV and search for the decay \psipptophietap}

\author{
M.~Ablikim$^{1}$, M.~N.~Achasov$^{5,b}$, P.~Adlarson$^{74}$, X.~C.~Ai$^{80}$, R.~Aliberti$^{35}$, A.~Amoroso$^{73A,73C}$, M.~R.~An$^{39}$, Q.~An$^{70,57}$, Y.~Bai$^{56}$, O.~Bakina$^{36}$
% ...
\\
\vspace{0.2cm}
(BESIII Collaboration)\\
\vspace{0.2cm} {\it
$^{1}$ Institute of High Energy Physics, Beijing 100049, People's Republic of China\\
% ...
$^{5}$ Budker Institute of Nuclear Physics SB RAS (BINP), Novosibirsk 630090, Russia\\
% ...
$^{35}$ Johannes Gutenberg University of Mainz, Johann-Joachim-Becher-Weg 45, D-55099 Mainz, Germany\\
$^{36}$ Joint Institute for Nuclear Research, 141980 Dubna, Moscow region, Russia\\
% ...
$^{39}$ Liaoning Normal University, Dalian 116029, People's Republic of China\\
% ...
$^{56}$ Southeast University, Nanjing 211100, People's Republic of China\\
$^{57}$ State Key Laboratory of Particle Detection and Electronics, Beijing 100049, Hefei 230026, People's Republic of China\\
% ...
$^{70}$ University of Science and Technology of China, Hefei 230026, People's Republic of China\\
% ...
$^{73}$ University of Turin and INFN, (A)University of Turin, I-10125, Turin, Italy; (B)University of Eastern Piedmont, I-15121, Alessandria, Italy; (C)INFN, I-10125, Turin, Italy\\
$^{74}$ Uppsala University, Box 516, SE-75120 Uppsala, Sweden\\
% ...
$^{80}$ Zhengzhou University, Zhengzhou 450001, People's Republic of China\\
\vspace{0.2cm}
% ...
$^{b}$ Also at the Novosibirsk State University, Novosibirsk, 630090, Russia\\
% ...
}}

\date{\today}

\begin{abstract}
The cross sections of the~\eetophietap~process at center-of-mass energies from 3.508 to 4.951~GeV are measured with high precision  using 26.1~fb$^{-1}$ data collected with the BESIII detector operating at the BEPCII storage ring. The cross sections are of the order of a few picobarn, and decrease as the center-of-mass energy increases as $s^{-n/2}$ with $n=4.35\pm 0.14$.
This result is in agreement with the Nambu-Jona-Lasinio model prediction of $n=3.5\pm 0.9$.
In addition, the charmless decay $\pspp\to\phi\etap$ is searched for by fitting the measured cross sections, yet no significant signal is observed. The upper limit of $\BR(\psi(3770)\to\phi\etap)$~at the 90\% confidence level is determined to be $2.3\times 10^{-5}$.

\end{abstract}

\pacs{Valid PACS appear here}

\maketitle

\section{Introduction}
The study of $\EE$ annihilation into light hadrons is essential to understand the production mechanism of light hadrons and reveal fundamental aspects of the strong interactions of light quarks~\cite{Davier:2013vna}.
Within the chiral perturbation framework, the Nambu-Jona-Lasinio (NJL) model has been successful in describing the dynamics of strong interactions between elementary particles and predicting the cross sections of the $\EE\to$ vector meson + pseudoscalar meson (VP) process over a range of center-of-mass (c.m.) energies ($\sqrt{s}$) from 2.3 to 5.5~GeV~\cite{Nambu:1961tp, Volkov:1986zb, Volkov:2005kw, Prades:1993ys}, here we only refer to the light hadrons final states.
Within large experimental uncertainties, the cross sections of the \eetorhoetap~process
measured by the \babar\ Collaboration~\cite{BaBar:2007qju} are consistent with the NJL model predictions \cite{Bystritskiy:2008pr}.
The NJL model also predicts that the cross sections of the \eetophietap~process from $\sqrt{s}$=3.5 to 5.0~GeV
are expected to to be $\sim 60$ picobarn and follow an asymptotic power-law behavior of $\sigma\propto s^{-n/2}$ with $n=3.5\pm 0.9$
as can be derived from the result in \cite{Bystritskiy:2008pr}. However, these predictions have never been tested with high-precision experimental data.

The $\pspp$, observed in the 1970s~\cite{int30}, is the first charmonium resonance above the $D\bar{D}$ production threshold. It is expected to decay almost entirely to Okubo-Zweig-Iizuka-allowed $D\bar{D}$ final states while the fractions of hadronic and radiative transitions to lower lying charmonium states and decays into light hadrons (states with $u$, $d$, $s$ quarks) are predicted to be small~\cite{Eichten:1974af, Eichten:1979ms}. Unexpectedly, the BES experiment measured a branching fraction for $\pspp\to{\rm non}$-$D\bar{D}$ to be $(14.7\pm 3.2)\%$ ~\cite{int37,int38,int39,int40}, while the CLEO result is $(-3.3\pm1.4^{+6.6}_{-4.8})\%$~\cite{int41}.
Taking the charmonium transition into account~\cite{int31,int32,int33,int34,int35}, a few percent of $\pspp$ decays into light hadrons are still allowed due to the large width of $\pspp$.
Although BES, CLEO-c, and BESIII searched for exclusive $\pspp$ decays into light hadrons, no significant signal for any decay was reported~\cite{int36, CLEO:2005tkm}. Until now, the sum of the branching fractions of the known ${\rm non}$-$D\bar{D}$ decays of the $\pspp$ remains to be less than 2\%~\cite{int1}.
Recently, BESIII measured the branching fraction for the inclusive decay $\psi(3770)\to J/\psi X$, which are consistent within error with the sum of published branching fractions $(0.47\pm0.06)\%$~\cite{int1}, of $\psi(3770)\to\pi^{+}\pi^{-}J/\psi, \pi^{0}\pi^{0}J/\psi, \eta J/\psi$, and $\gamma \chi_{cJ}$ with $J=0, 1, 2$~\cite{BESIII:2020xfc}.

Although the $\pspp$ is believed to be primarily the $1^{3}D_{1}$ state of the $c\bar{c}$ system, its large leptonic width of $(0.262\pm0.018)$ keV indicates substantial mixing with the $S$-wave state~\cite{Eichten:1978tg, Kuang:1989ub}. The $S$- and $D$-wave mixing scheme has been used to explain the ``$\rho\pi$ puzzle" in $\psi(2S)$ and $J/\psi$ decays~\cite{int42, int421} and may suggest that the small branching fractions for $\psi(2S)\to$ VP decays can be enhanced in $\pspp\to$ VP decays, including $\rho\pi$, $\phi\eta$, $\phi\etap$, and so on.
Analogous to the observed charmless decay of $\pspp\to\phi\eta$~\cite{int36}, we search for the related decay $\pspp\to\phi\eta'$.
By using the known branching fractions of $J/\psi\to\phi\eta'$ and $\psi(2S)\to\phi\eta'$~\cite{int1}, and taking into account the mixing angle $\theta$ between $S$- and $D$-wave, the $\pspp\to\phi\etap$ decay branching fraction ($\BR_{\phi\etap}$) is predicted to be $(0.46\pm0.18)\times 10^{-5}<\BR_{\phi\etap}<(3.5\pm 1.3) \times10^{-5}$
~\cite{int45, int46}. This decay can be searched for with the BESIII data at the $\pspp$ peak and in its vicinity.

A recent BESIII paper~\cite{BESIII:2022tjc} reported a study of the \eetophietap~process and a search for the charmless decay $Y(4230)\to\phi\eta'$ with part of the data used in this analysis.
In this article, we report a measurement of the cross sections of the \eetophietap~process with 26.1~fb$^{-1}$ data collected at $\sqrt{s}=3.508\sim 4.951$~GeV~\cite{dec1} to test the NJL model prediction and to search for the charmless decay $\pspp\to \phi\etap$.
Due to the low integrated luminosity of the $\psi(3770)$ scan data samples, we do not use the $\psi(3770)$ scan presented in the BESIII work~\cite{BESIII:2020xfc}.

\section{Detector and data samples}\label{sec:dec}

The BESIII detector~\cite{dec1} records symmetric $e^+e^-$ collisions provided by the BEPCII storage ring~\cite{dec2} in the $\sqrt{s}$ range from 2.0 to 4.95~GeV, with a peak luminosity of $1.05\times 10^{33}~\text{cm}^{-2}\text{s}^{-1}$ achieved at $\sqrt{s} = 3.77~\text{GeV}$. The cylindrical core of BESIII detector consists of a helium-based multilayer drift chamber (MDC), a plastic scintillator time-of-flight system (TOF), and a CsI (Tl) electromagnetic calorimeter (EMC), which are all enclosed in a superconducting solenoidal magnet providing a 1.0~T magnetic field~\cite{detvis}. The solenoid is supported by an octagonal flux-return yoke with resistive plate counter muon identifier modules interleaved with steel. The acceptance of charged particles and photons is 93\% over 4$\pi$ solid angle. The charged particle momentum resolution at 1 GeV/$c$ is 0.5\%, and the specific ionization energy loss $dE/dx$ resolution is 6\% for the electrons from Bhabha scattering. The EMC measures photon energies with a resolution of 2.5\% (5\%) at 1~GeV in the barrel (end cap) region. The time resolution of the TOF barrel part is 68~ps, while that of the end cap part is 110~ps. The end cap TOF system was upgraded in 2015 with multi-gap resistive plate chamber technology, providing a time resolution of 60~ps~\cite{dec3, dec4}.

Details of the data samples used in this analysis are listed in Table~\ref{tab:DressedCS-0.5pi}.  The optimization of event selection criteria and efficiency determination are based on Monte Carlo (MC) simulations. The {\sc geant4}-based~\cite{geant4} simulation software, BESIII Object Oriented Simulation Tool ({\sc boost})~\cite{boost}, includes the geometric description of the BESIII detectors.
The beam energy spread and initial state radiation (ISR) in the $\EE$ annihilation are modeled with the generator {\sc kkmc}~\cite{kkmc1, kkmc2}.

Samples of MC events for the signal process $\EE\to\phi\etap$ are generated using the HELAMP model within the {\sc evtgen} framework~\cite{evtgen1, evtgen2}, which allow simulations of any two-body decay by specifying the helicity amplitudes for the final state particles.
The subsequent decay \phitokk~is generated with the Vector-Scalar-Scalar (VSS) model in which a vector particle decays into two scalars.
The angular distribution of
$K$ is proportional to sin$^{2}\theta_{K}$ in the $\phi$ helicity frame~\cite{evtgen1}, where $\theta_{K}$ is the helicity angle of $K$ in $\phi$ helicity frame.
The decay \etaptogammapipi~is generated with the DIY generator taking into account both $\rho-\omega$ interference and the box anomaly~\cite{Qin:2017vkw}.
The decays \etaptoetapipi and \etatogammagamma~are both generated uniformly in phase space. Final state radiation associated with charged final state particles is handled by {\sc photos}~\cite{photos}.

The inclusive MC samples generated at $\sqrt{s}$=3.650, 3.773, 4.178, 4.416, and 4.843~GeV are used to estimate the possible background. The continuum processes are incorporated in {\sc kkmc}, and the QED processes such as Bhabha scattering,~\MM, \TT, and \GG~events are generated with {\sc kkmc} and {\sc babayaga}~\cite{Babayaga}.
All particle decays are modelled with {\sc evtgen} using branching fractions either taken from
the Particle Data Group (PDG)~\cite{int1}, when available, or otherwise estimated with {\sc lundcharm} for $J/\psi$ and $\psi(2S)$ decays~\cite{lund1, lund2}.

\section{Event Selection}\label{sec:event}
Charged tracks detected in the MDC are required to be within a polar angle ($\theta$) range of $|\cos\theta|<0.93$,
where $\theta$ is defined with respect to the $z$-axis, the symmetry axis of the MDC. For charged tracks,
the distances of the closest approach to the interaction point along the $z$-axis
and in the transverse plane must be less than 10~cm
and 1~cm, respectively.
\begin{figure*}[htbp]
  \centering
  \subfigure[]{
  \label{Fig1.sub.1}
  \includegraphics[width=0.45\textwidth]{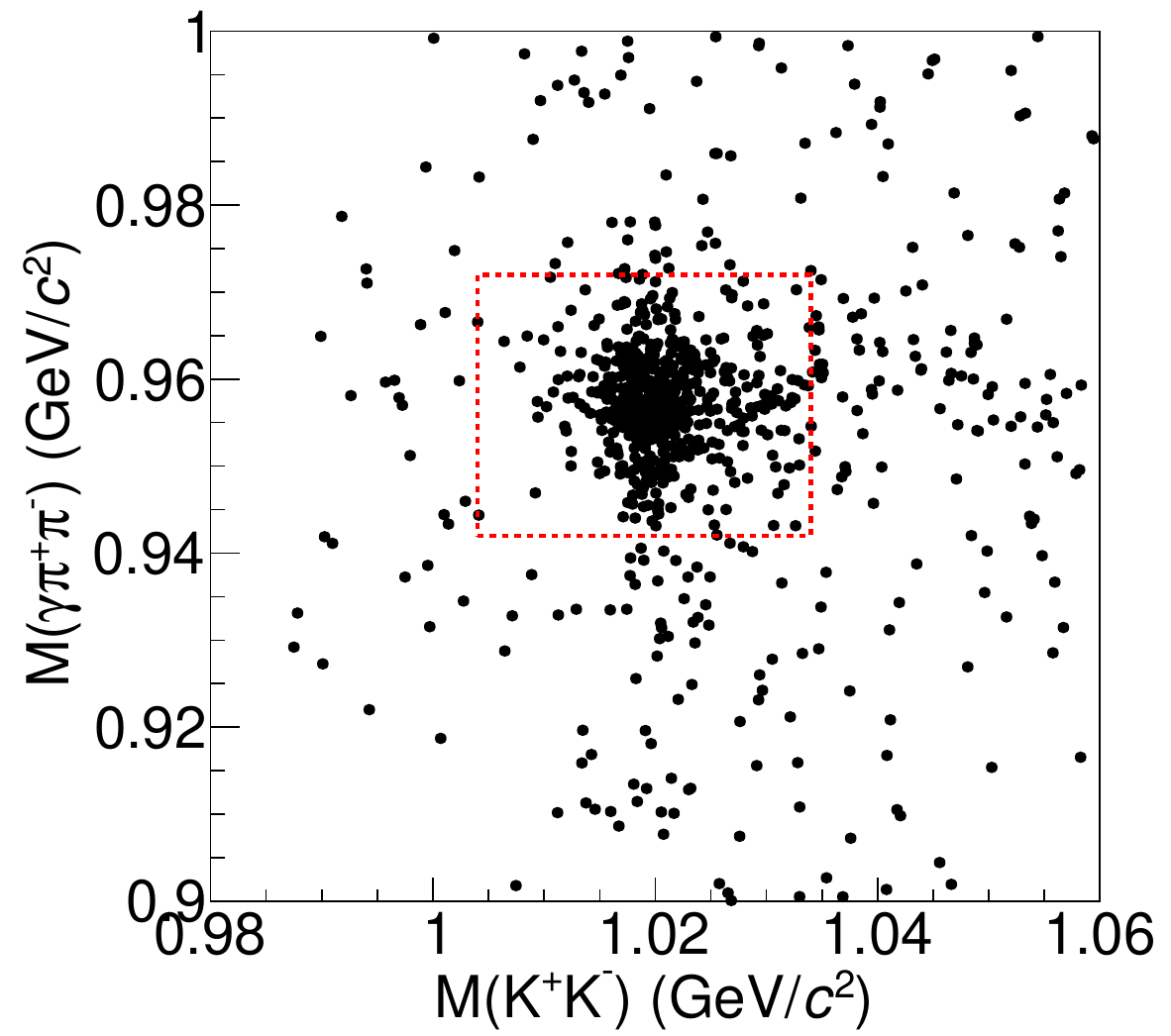}}
  \subfigure[]{
  \label{Fig1.sub.2}
  \includegraphics[width=0.45\textwidth]{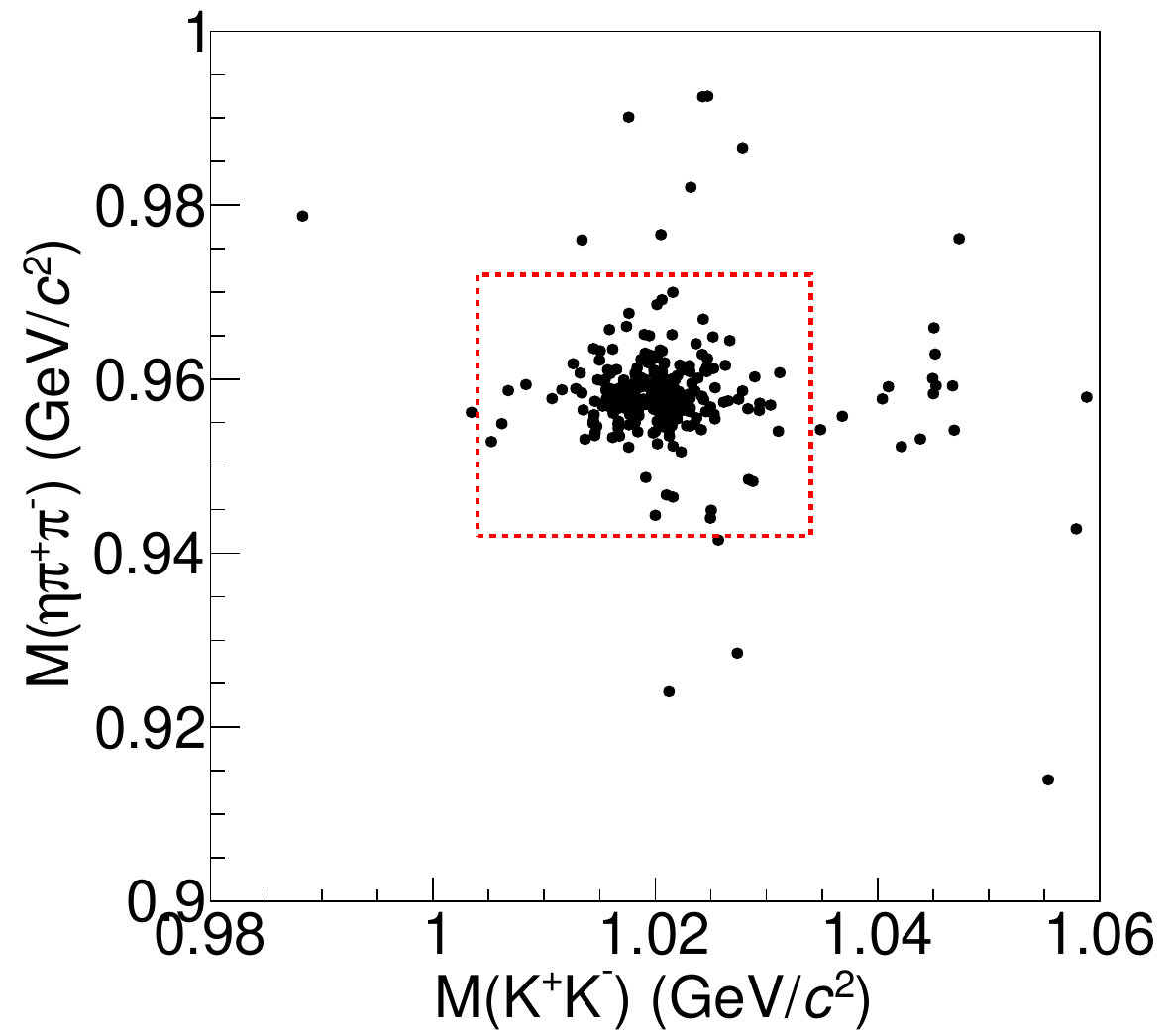}}
\caption{The two-dimensional distribution of $M(\KpKm)$ versus $M(\gampipi)$ for mode~I (a) and versus $M(\etapipi)$ for mode~II (b) from data at $\sqrt s =$3.773~GeV. The red rectangle indicates the signal region.}\label{pic:Mphi-Metep}
\end{figure*}
Photon candidates are identified using showers in the EMC. The deposited energy of each shower must be more than 25~MeV in the barrel region ($|\cos \theta|< 0.80$) and more than 50~MeV in the end cap region ($0.86 <|\cos \theta|< 0.92$). To significantly reduce showers that originate from charged tracks, the angle subtended by the EMC shower and the position of the closest charged track at the EMC must be greater than 10 degrees as measured from the interaction point. To suppress electronic noise and showers unrelated to the event, the difference between the EMC time and the event start time is required to be within [0,~700]~ns.
\begin{figure*}[ht]
  \centering
  \subfigure[]{
  \label{Fig2.sub.1}
  \includegraphics[width=0.45\textwidth]{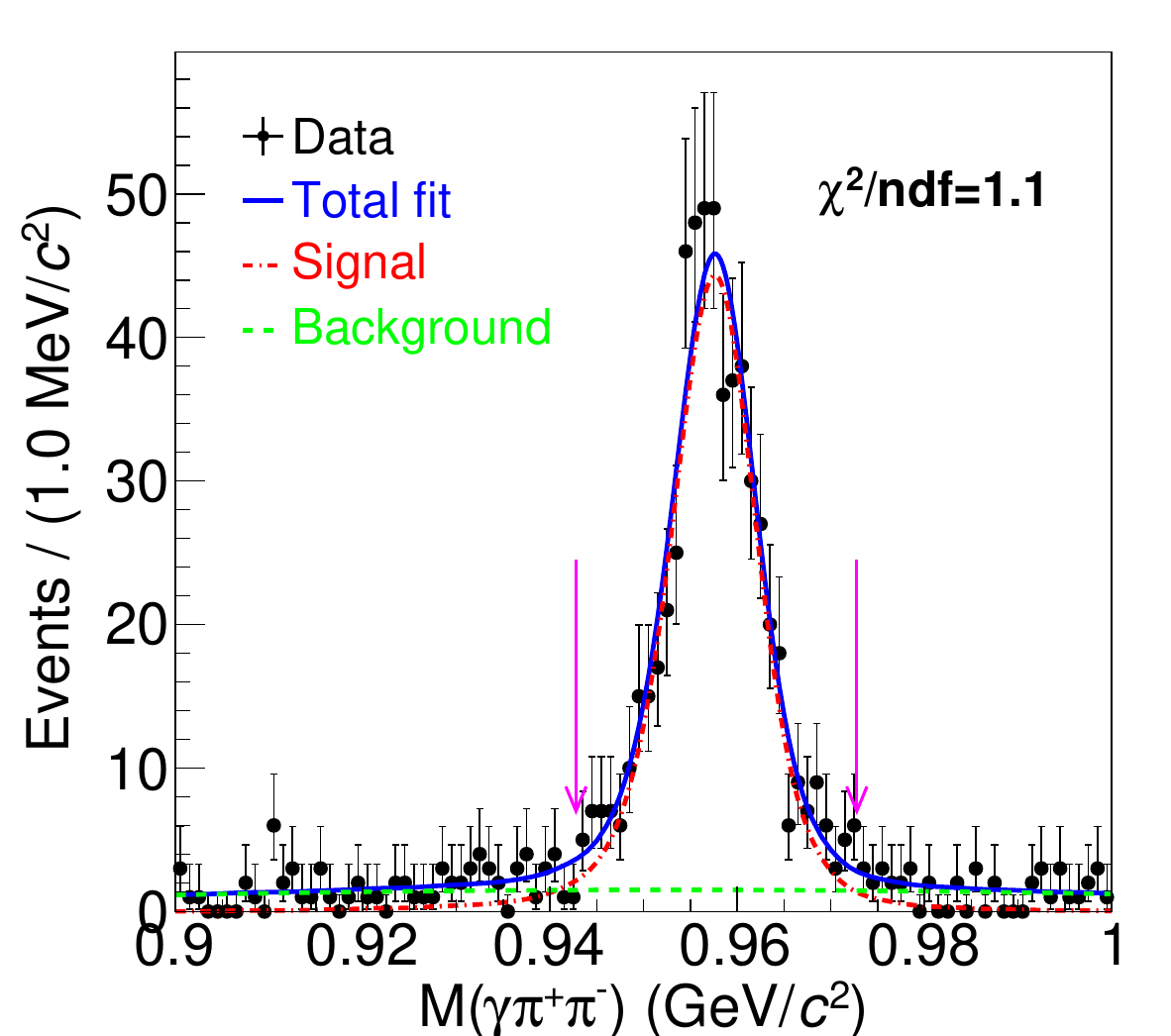}}
  \subfigure[]{
  \label{Fig2.sub.2}
   \includegraphics[width=0.45\textwidth]{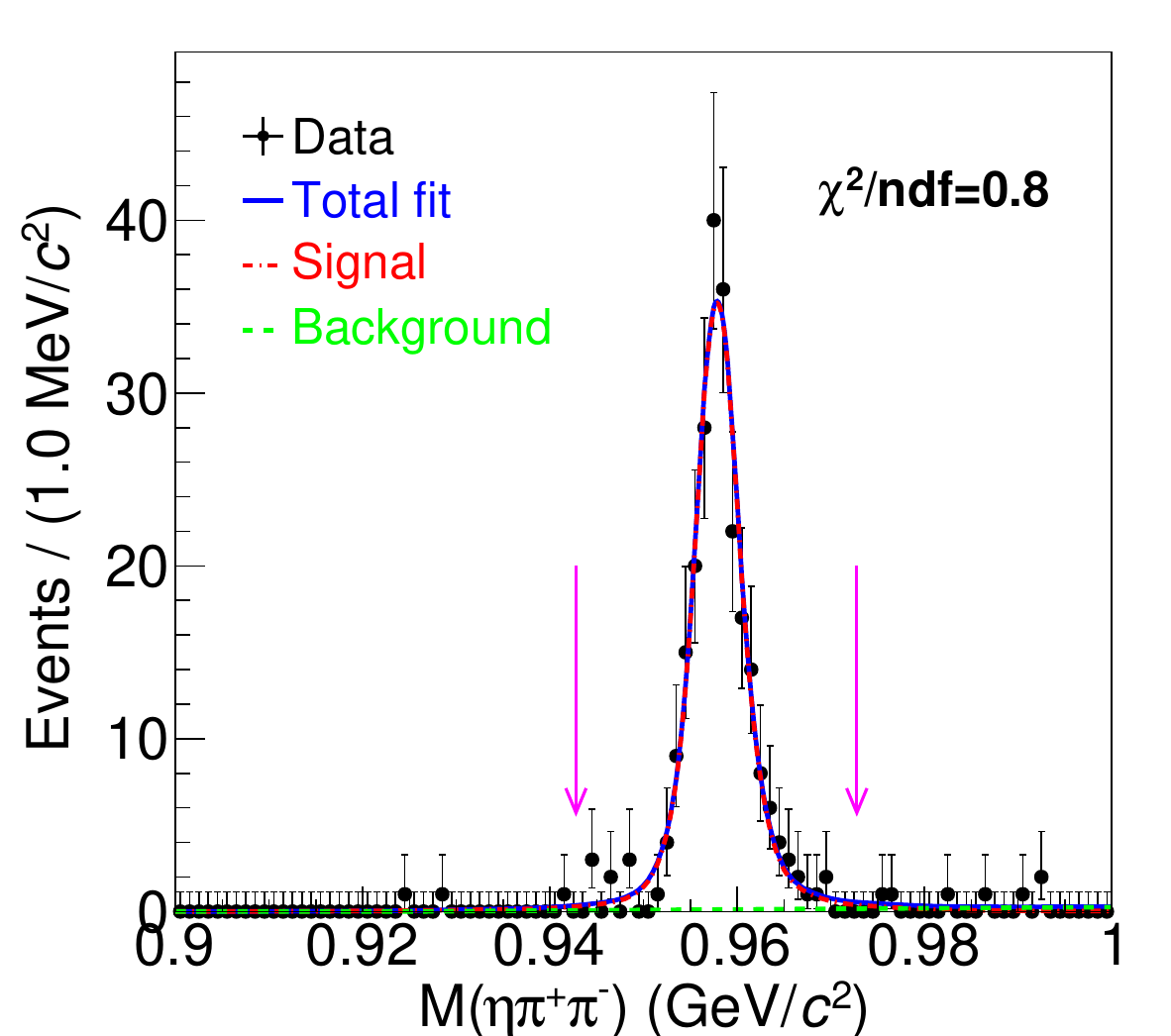}}
\caption{The unbinned maximum likelihood fits on the distributions of $M(\gampipi)$ for mode I (a) and $M(\etapipi)$ for mode II (b) at $\sqrt s =$3.773~GeV. The black dots with error bars represent the data sample, the red dot-dashed lines represent the signal MC sample, the green dashed lines are the fitted background and the blue solid lines represent the total fit. The signal region lies between the two pink solid arrows.}\label{pic:3773-fit}
\end{figure*}
To reconstruct \eetophietap, the $\phi$ candidate is reconstructed via the $\KpKm$ decays, and the $\etap$ candidate is reconstructed
via $\gampipi$~(mode I) and $\etapipi$~with \etatogammagamma~(mode II).
We require exactly four charged tracks with zero net-charge and at least one (two) photon(s) candidates in an event for mode I~(II). A vertex fit is performed on the four charged tracks to ensure they originate from the same vertex.
In order to select $\phi$ candidates, for both modes I and II, we test all possible combinations of oppositely charged particles
which are assumed to be kaons. A $\phi$ candidate is selected if it satisfies $|M(\KpKm) - m_{\phi}| < 0.04$~GeV/$c^{2}$ ($M(\KpKm)$ is the invariant mass of $K^+K^-$ and $m_{\phi}$ is the nominal mass of $\phi$ from PDG~\cite{int1}).
To remove a combinatorial background, events with more than one selected $\phi$ candidate are rejected. According to the MC simulation, this reduces the signal statistics by about 1\%.
A four-constraint (4C) kinematic fit, which constrains the sum of four momentum of the final-state particles to the four momentum of the initial colliding beams, is performed for the final state particles of mode I in each event to improve resolutions and suppress the background.
If there is more than one combination due to multiple photons in an event, the one with the smallest $\chi^2_{\rm 4C}$ is selected. The $\chi^2_{\rm 4C} $ is required to be less than 100 for mode~I.
For mode~II, a five-constraint (5C) kinematic fit is performed with an additional constraint to the invariant mass of the $\eta$ candidate to the $\eta$ nominal mass from PDG~\cite{int1}.
We loop over all possible combinations of $\eta$ candidates and select the one with the least $\chi^2_{\rm 5C} $ of the
kinematic fit. The $\chi^2_{\rm 5C} $ is required to be less than 200 for mode~II.
All the selection criteria have been optimized by maximizing the
figure of merit $S/\sqrt{S+B}$, where $S(B)$ is the number of signal (background) events in the signal region, determined by MC simulation.
The signal candidates for $\phi$ and $\etap$ mesons are required to be within the mass ranges $1.004\leq M(\KpKm)\leq 1.034$ GeV/$c^{2}$ for both modes, and $0.943 \leq M(\gamma(\eta)\pi^+\pi^-) \leq 0.973$ GeV/$c^{2}$ for mode~I~(II).

After applying the mentioned selection criteria, Fig.~\ref{pic:Mphi-Metep} shows the scatter plots of $M(\KpKm)$ versus $M(\gampipi)$ and $M(\etapipi)$ in data at $\sqrt s =$3.773~GeV for modes I and II. Two clear accumulations of signal are observed in Fig.~\ref{pic:Mphi-Metep} around the intersection of the $\phi$ and $\etap$ signal regions. We have used the sidebands of $\phi$ and $\etap$ in data to check the possible peaking backgrounds from $K^+K^-\etap$ and $\gamma(\eta)\pi^+\pi^-\phi$, and we do not observe obvious peaking backgrounds.

To extract the signal yield, an unbinned maximum likelihood fit is performed on the $M(\gampipi)$ and $M(\etapipi)$ distributions for modes~I and~II, respectively. In the fit, the signal is described by the line shape of the $\eta'$ from MC simulation convolved with a Gaussian function, which accounts for the difference in resolutions between data and MC simulation, and the background is described by a linear function. The parameters of the Gaussian function and the linear function are free in the fit.
Due to the low statistics at $\sqrt s=$3.670, 3.867, 3.871, 4.278, 4.308, 4.467, 4.527, 4.699, 4.740, 4.918, and 4.951~GeV, the parameters of the Gaussian function are fixed to the values obtained from the nearby data samples with large statistics for mode~I.
For mode~II, the background level is very low and the parameters of the Gaussian function are free in the fit. The fit results at $\sqrt s=$3.773~GeV for modes~I and~II are shown in Fig.~\ref{pic:3773-fit}. The results for the number of events within the signal region, $N^{\rm signal}$, are obtained from the fits for all data samples and listed in Table~\ref{tab:DressedCS-0.5pi}. The uncertainty of $N^{\rm signal}$ obtained from the fit
may be smaller than $\sqrt{N^{\rm signal}}$, because the fit range is larger than the signal region. The uncertainty of $N^{\rm signal}$ is estimated with MINOS in TMINUIT~\cite{sta}.
\begin{table*}[htbp]
\caption{Signal yields and cross sections for the \eetophietap~process at different c.m. energies ($\sqrt s$)~\cite{BESIII:2015zbz,BESIII:2015qfd,BESIII:2022xyz}. Here, ${\cal L}$, ${N^{\rm signal}}$, $\epsilon$, $(1+\delta)$, and $\frac{1}{{|1-\Pi|}^2}$ are the integrated luminosity~\cite{BESIII:2015qfd,BESIII:2022xyz,BESIII:2022xyz2}, the number of signal events in the signal region, the detection efficiency, the radiative correction factor, and the vacuum polarization factor, respectively.
$\sigma^{\rm dressed}_{\rm com}$ and $\sigma^{\rm Born}_{\rm com}$ are the combined dressed cross section and Born cross section of the \eetophietap~process, where the first uncertainties are statistical, and the second uncertainties are systematic. The subscripts I and II represent modes I and II, respectively.}\label{tab:DressedCS-0.5pi}
\centering
\normalsize
\begin{tabular}{cr@{.}lr@{.}lr@{.}lcccccc}
\toprule
\hline\hline
$\sqrt s$ (GeV)  &\multicolumn{2}{c}{${\cal L}$ $({\rm pb^{-1}})$}   &\multicolumn{2}{c}{${N_{\rm I}^{\rm signal}}$}   &\multicolumn{2}{c}{${N_{\rm II}^{\rm signal}}$}  &$\epsilon_{\rm I}$ (\%)  &$\epsilon_{\rm II}$ (\%)  &$(1+\delta)$ &$\frac{1}{{|1-\Pi|}^2}$   &$\sigma^{\rm dressed}_{\rm com} ({\rm pb})$   &$\sigma^{\rm Born}_{\rm com} ({\rm pb})$  \\\hline
\midrule
3.508     &181 &8    &46 &1$^{+7.3}_{-7.3}$      &15 &6$^{+4.3}_{-3.6}$      &32.52\%        &25.79\%        &0.734      &1.044      &6.75$^{+0.93}_{-0.89}\pm$0.34  &6.47$^{+0.89}_{-0.85}\pm$0.32  \\
3.511     &184 &6    &50 &3$^{+7.5}_{-7.5}$      &22 &4$^{+5.4}_{-4.7}$      &32.47\%        &26.02\%        &0.737      &1.044      &7.79$^{+0.99}_{-0.95}\pm$0.39   &7.46$^{+0.95}_{-0.91}\pm$0.37  \\
3.582     &85  &7    &25 &0$^{+5.2}_{-5.2}$      &7  &5$^{+3.1}_{-2.5}$      &31.09\%        &25.02\%        &0.792      &1.039      &7.28$^{+1.36}_{-1.29}\pm$0.36    &7.01$^{+1.31}_{-1.24}\pm$0.35  \\
3.650     &453 &9    &94 &0$^{+10.4}_{-10.4}$    &37 &5$^{+6.7}_{-6.0}$      &30.48\%        &24.23\%        &0.915      &1.021      &4.93$^{+0.46}_{-0.45}\pm$0.25    &4.83$^{+0.45}_{-0.44}\pm$0.24  \\
3.670     &84  &7    &13 &2$^{+3.6}_{-3.6}$      &6  &8$^{+2.9}_{-2.2}$      &30.12\%        &24.20\%        &0.839      &0.994      &4.42$^{+1.02}_{-0.93}\pm$0.22    &4.45$^{+1.03}_{-0.94}\pm$0.24  \\
3.773     &2931&8    &538&8$^{+24.9}_{-24.9}$    &237&0$^{+15.6}_{-15.6}$      &28.33\%        &22.28\%        &0.960      &1.056      &4.63$^{+0.18}_{-0.18}\pm$0.23    &4.39$^{+0.17}_{-0.17}\pm$0.22  \\
3.808     &50  &5    &7  &4$^{+3.2}_{-2.6}$      &2  &9$^{+2.0}_{-1.4}$      &30.09\%        &23.88\%        &0.896      &1.056      &3.59$^{+1.32}_{-1.03}\pm$0.18    &3.40$^{+1.25}_{-0.98}\pm$0.17  \\
3.867     &108 &9    &8  &8$^{+3.7}_{-3.1}$      &6  &8$^{+3.1}_{-2.5}$      &29.26\%        &23.45\%        &0.918      &1.051      &2.53$^{+0.78}_{-0.64}\pm$0.13    &2.40$^{+0.74}_{-0.61}\pm$0.12  \\
3.871     &110 &3    &19 &9$^{+4.6}_{-4.6}$      &5  &8$^{+2.7}_{-2.1}$      &29.29\%        &23.17\%        &0.919      &1.051      &4.12$^{+0.85}_{-0.81}\pm$0.21    &3.92$^{+0.81}_{-0.77}\pm$0.20  \\
3.896     &52  &6    &11 &0$^{+3.8}_{-3.2}$      &3  &9$^{+2.3}_{-1.6}$      &29.44\%        &23.61\%        &0.927      &1.049      &4.91$^{+1.46}_{-1.18}\pm$0.25    &4.68$^{+1.40}_{-1.12}\pm$0.23  \\
4.008     &482 &0    &52 &5$^{+7.9}_{-7.9}$      &30 &4$^{+6.0}_{-5.4}$      &28.52\%        &22.79\%        &0.964      &1.044      &2.96$^{+0.35}_{-0.34}\pm$0.15    &2.84$^{+0.34}_{-0.33}\pm$0.14  \\
4.085     &52  &9    &7  &1$^{+3.1}_{-2.5}$      &2  &0$^{+1.7}_{-1.1}$      &28.34\%        &22.61\%        &0.988      &1.051      &2.91$^{+1.13}_{-0.87}\pm$0.15    &2.77$^{+1.08}_{-0.83}\pm$0.14  \\
4.128     &401 &5    &37 &4$^{+6.5}_{-6.5}$      &21 &4$^{+4.9}_{-4.2}$      &27.66\%        &21.95\%        &1.047      &1.052      &2.40$^{+0.33}_{-0.32}\pm$0.12    &2.28$^{+0.32}_{-0.30}\pm$0.11  \\
4.157     &408 &7    &41 &3$^{+6.8}_{-6.8}$      &9  &2$^{+3.6}_{-2.9}$      &27.75\%        &22.26\%        &1.052      &1.053      &2.00$^{+0.31}_{-0.29}\pm$0.10    &1.90$^{+0.29}_{-0.28}\pm$0.10  \\
4.178     &3189&0    &299&1$^{+17.9}_{-17.9}$    &113&6$^{+11.1}_{-10.4}$      &27.82\%        &21.80\%        &1.014      &1.054      &2.19$^{+0.11}_{-0.11}\pm$0.11    &2.08$^{+0.11}_{-0.10}\pm$0.10  \\
4.189     &570 &0    &63 &5$^{+8.3}_{-8.3}$      &21 &3$^{+5.2}_{-4.5}$      &27.61\%        &21.93\%        &1.017      &1.056      &2.52$^{+0.29}_{-0.28}\pm$0.13    &2.38$^{+0.28}_{-0.27}\pm$0.12  \\
4.199     &526 &0    &36 &0$^{+6.4}_{-6.4}$      &17 &6$^{+4.9}_{-4.1}$      &27.92\%        &21.99\%        &1.020      &1.056      &1.70$^{+0.26}_{-0.24}\pm$0.09    &1.61$^{+0.24}_{-0.23}\pm$0.08  \\
4.209     &572 &0    &46 &5$^{+7.3}_{-7.3}$      &10 &7$^{+3.6}_{-2.9}$      &27.45\%        &21.63\%        &1.021      &1.057      &1.70$^{+0.24}_{-0.23}\pm$0.08    &1.61$^{+0.23}_{-0.22}\pm$0.08  \\
4.219     &569 &2    &48 &0$^{+7.1}_{-7.1}$      &18 &7$^{+4.8}_{-4.1}$      &27.43\%        &21.65\%        &1.024      &1.056      &1.98$^{+0.26}_{-0.24}\pm$0.10    &1.88$^{+0.24}_{-0.23}\pm$0.09  \\
4.226     &1100&9    &106&2$^{+10.9}_{-10.9}$    &53 &6$^{+7.6}_{-6.9}$      &27.90\%        &22.14\%        &1.026      &1.056      &2.41$^{+0.20}_{-0.19}\pm$0.12    &2.28$^{+0.19}_{-0.18}\pm$0.11  \\
4.236     &530 &3    &44 &1$^{+6.9}_{-6.9}$      &17 &0$^{+4.5}_{-3.9}$      &27.57\%        &21.99\%        &1.029      &1.056      &1.93$^{+0.26}_{-0.25}\pm$0.10    &1.82$^{+0.25}_{-0.24}\pm$0.09  \\
4.244     &538 &1    &45 &3$^{+7.1}_{-7.1}$      &18 &5$^{+4.6}_{-3.9}$      &27.78\%        &22.04\%        &1.031      &1.056      &1.97$^{+0.26}_{-0.25}\pm$0.10    &1.86$^{+0.25}_{-0.24}\pm$0.09  \\
4.258     &828 &4    &73 &5$^{+9.0}_{-9.0}$      &25 &1$^{+6.1}_{-4.8}$      &27.82\%        &22.04\%        &1.035      &1.054      &1.96$^{+0.22}_{-0.20}\pm$0.10    &1.86$^{+0.21}_{-0.19}\pm$0.09  \\
4.267     &531 &1    &45 &2$^{+6.9}_{-6.9}$      &23 &4$^{+5.5}_{-4.8}$      &27.76\%        &21.95\%        &1.038      &1.053      &2.13$^{+0.27}_{-0.26}\pm$0.11    &2.02$^{+0.26}_{-0.25}\pm$0.10  \\
4.278     &175 &7    &7  &9$^{+3.2}_{-2.6}$      &4  &9$^{+2.5}_{-1.9}$      &27.19\%        &21.48\%        &1.042      &1.053      &1.22$^{+0.39}_{-0.31}\pm$0.06    &1.16$^{+0.37}_{-0.29}\pm$0.06  \\
4.288     &502 &4    &32 &0$^{+6.3}_{-6.3}$      &15 &6$^{+4.2}_{-3.6}$      &26.99\%        &21.29\%        &1.083      &1.053      &1.54$^{+0.25}_{-0.24}\pm$0.08    &1.46$^{+0.23}_{-0.22}\pm$0.07  \\
4.308     &45  &1    &1  &9$^{+1.7}_{-1.0}$      &1  &0$^{+1.3}_{-0.0}$      &27.53\%        &21.83\%        &1.051      &1.052      &1.06$^{+0.78}_{-0.36}\pm$0.05    &1.00$^{+0.74}_{-0.35}\pm$0.05  \\
4.312     &501 &2    &34 &0$^{+6.1}_{-6.1}$      &12 &7$^{+4.0}_{-3.4}$      &27.08\%        &21.06\%        &1.087      &1.052      &1.51$^{+0.24}_{-0.23}\pm$0.08    &1.44$^{+0.22}_{-0.22}\pm$0.07  \\
4.337     &505 &0    &27 &1$^{+5.6}_{-5.6}$      &12 &7$^{+3.8}_{-3.2}$      &26.98\%        &21.27\%        &1.093      &1.051      &1.27$^{+0.22}_{-0.21}\pm$0.06    &1.21$^{+0.21}_{-0.20}\pm$0.06  \\
4.358     &543 &9    &52 &7$^{+7.4}_{-7.4}$      &12 &5$^{+4.1}_{-3.4}$      &27.69\%        &21.65\%        &1.065      &1.051      &1.94$^{+0.25}_{-0.24}\pm$0.10    &1.85$^{+0.24}_{-0.23}\pm$0.09  \\
4.377     &522 &7    &24 &7$^{+5.3}_{-5.3}$      &17 &6$^{+4.4}_{-3.9}$      &26.92\%        &21.13\%        &1.102      &1.051      &1.30$^{+0.21}_{-0.20}\pm$0.07    &1.24$^{+0.20}_{-0.19}\pm$0.06  \\
4.396     &507 &8    &37 &5$^{+6.3}_{-6.3}$      &10 &6$^{+3.8}_{-3.2}$      &26.80\%        &20.86\%        &1.106      &1.051      &1.53$^{+0.23}_{-0.22}\pm$0.08    &1.45$^{+0.22}_{-0.21}\pm$0.07  \\
4.416     &1090&7    &83 &3$^{+9.4}_{-9.4}$      &24 &4$^{+5.2}_{-4.6}$      &27.14\%        &21.32\%        &1.082      &1.052      &1.60$^{+0.16}_{-0.16}\pm$0.08    &1.52$^{+0.15}_{-0.15}\pm$0.08  \\
4.436     &569 &9    &47 &0$^{+7.5}_{-7.5}$      &14 &4$^{+4.3}_{-3.6}$      &26.65\%        &21.00\%        &1.116      &1.054      &1.72$^{+0.24}_{-0.23}\pm$0.09    &1.64$^{+0.23}_{-0.22}\pm$0.08  \\
4.467     &111 &1    &4  &8$^{+2.5}_{-1.9}$      &1  &0$^{+1.3}_{-0.7}$      &26.94\%        &20.93\%        &1.096      &1.055      &0.85$^{+0.41}_{-0.30}\pm$0.04    &0.80$^{+0.39}_{-0.28}\pm$0.04  \\
4.527     &112 &1    &6  &6$^{+2.8}_{-2.2}$      &2  &9$^{+2.0}_{-1.4}$      &26.64\%        &20.61\%        &1.112      &1.054      &1.37$^{+0.50}_{-0.38}\pm$0.07    &1.30$^{+0.47}_{-0.36}\pm$0.06  \\
4.600     &586 &9    &24 &0$^{+4.9}_{-4.9}$      &14 &6$^{+4.1}_{-3.5}$      &26.40\%        &20.54\%        &1.132      &1.055      &1.05$^{+0.17}_{-0.16}\pm$0.05    &1.00$^{+0.17}_{-0.16}\pm$0.05  \\
4.612     &103 &8    &4  &9$^{+2.5}_{-1.9}$      &0  &0$^{+0.5}_{-0.0}$      &26.01\%        &19.95\%        &1.158      &1.055      &0.75$^{+0.39}_{-0.29}\pm$0.04    &0.71$^{+0.37}_{-0.28}\pm$0.04  \\
4.628     &521 &5    &28 &5$^{+5.6}_{-5.6}$      &8  &7$^{+3.5}_{-2.8}$      &25.85\%        &19.76\%        &1.162      &1.054      &1.14$^{+0.20}_{-0.19}\pm$0.06    &1.08$^{+0.19}_{-0.18}\pm$0.05  \\
4.641     &552 &4    &27 &5$^{+5.4}_{-5.4}$      &8  &8$^{+3.6}_{-2.9}$      &25.90\%        &19.74\%        &1.164      &1.054      &1.05$^{+0.19}_{-0.18}\pm$0.05    &0.99$^{+0.18}_{-0.17}\pm$0.05  \\
4.661     &529 &6    &27 &6$^{+5.1}_{-5.1}$      &17 &5$^{+4.4}_{-3.8}$      &25.59\%        &19.59\%        &1.169      &1.054      &1.37$^{+0.20}_{-0.19}\pm$0.07    &1.30$^{+0.19}_{-0.18}\pm$0.06  \\
4.682     &1669&3    &66 &4$^{+8.5}_{-8.5}$      &21 &2$^{+5.3}_{-4.4}$      &25.69\%        &19.53\%        &1.174      &1.054      &0.84$^{+0.10}_{-0.09}\pm$0.04    &0.79$^{+0.09}_{-0.09}\pm$0.04  \\
4.699     &536 &5    &16 &7$^{+4.3}_{-4.3}$      &8  &5$^{+3.3}_{-2.7}$      &25.76\%        &19.25\%        &1.178      &1.055      &0.75$^{+0.16}_{-0.15}\pm$0.04    &0.71$^{+0.15}_{-0.14}\pm$0.04  \\
4.740     &164 &3    &11 &3$^{+3.4}_{-3.4}$      &2  &9$^{+2.0}_{-1.4}$      &26.18\%        &19.52\%        &1.188      &1.055      &1.35$^{+0.37}_{-0.35}\pm$0.07    &1.28$^{+0.35}_{-0.33}\pm$0.06  \\
4.750     &367 &2    &8  &9$^{+3.6}_{-3.0}$      &8  &8$^{+3.3}_{-2.6}$      &26.23\%        &19.61\%        &1.192      &1.055      &0.75$^{+0.21}_{-0.17}\pm$0.04    &0.71$^{+0.20}_{-0.16}\pm$0.04  \\
4.781     &512 &8    &20 &2$^{+4.7}_{-4.7}$      &11 &7$^{+3.7}_{-3.1}$      &25.82\%        &19.57\%        &1.199      &1.055      &0.97$^{+0.18}_{-0.17}\pm$0.05    &0.92$^{+0.17}_{-0.16}\pm$0.05  \\
4.843     &527 &3    &25 &7$^{+5.3}_{-5.3}$      &7  &5$^{+3.2}_{-2.5}$      &25.63\%        &19.13\%        &1.213      &1.056      &0.98$^{+0.18}_{-0.17}\pm$0.05    &0.93$^{+0.17}_{-0.16}\pm$0.05  \\
4.918     &208 &1    &9  &4$^{+3.3}_{-2.7}$      &1  &9$^{+1.7}_{-1.1}$      &25.04\%        &18.94\%        &1.234      &1.056      &0.85$^{+0.28}_{-0.22}\pm$0.04    &0.80$^{+0.26}_{-0.21}\pm$0.04  \\
4.951     &160 &4    &6  &4$^{+2.8}_{-2.2}$      &4  &9$^{+2.8}_{-2.2}$      &24.78\%        &18.49\%        &1.244      &1.056      &1.11$^{+0.39}_{-0.30}\pm$0.06    &1.05$^{+0.37}_{-0.29}\pm$0.05  \\
\hline\hline
\bottomrule
\end{tabular}
\end{table*}
\section{CROSS SECTION MEASUREMENT OF THE \eetophietap~PROCESS}\label{sec:cs}
The dressed cross section of the \eetophietap~process at each energy point is calculated with
 \begin{equation}\label{equ:42}
 \sigma^{\rm dressed}_i = \frac{\sigma^{\rm Born}_i}{{|1-\Pi|}^2}= \frac {N^{\rm signal}_i} {{{\cal L} \cdot (1+\delta)\cdot \epsilon_i \cdot\BR_i}},
 \end{equation}
where $\sigma^{\rm Born}_i$ is the Born cross section, $N^{\rm signal}_i$ is the number of signal events extracted from the fit to the $M(\gamma\pi^+\pi^-)$ or $M(\etapipi)$ distribution as described in Sec.~\ref{sec:event}, ${\cal L}$ is the integrated luminosity, $\epsilon_i$ is the selection efficiency obtained from the MC simulation, and $\BR_i$ is the product of the branching fractions $\BR(\phi\to K^{+}K^{-})\cdot \BR(\etap\to\gamma\pi^{+}\pi^{-})$ or $\BR(\phi\to K^{+}K^{-})\cdot \BR(\etap\to\eta\pi^{+}\pi^{-})\cdot\BR$(\etatogammagamma), taken from PDG \cite{int1}. The subscript $i$ represents mode~I or mode~II. The vacuum polarization factor $\frac{1}{{|1-\Pi|}^2}$ is taken from Ref.~\cite{WorkingGrouponRadiativeCorrections:2010bjp}, and the radiative correction factor $(1+\delta)$ is  defined as
\begin{equation}\label{equ:43}
 {(1+\delta)  = \frac {\sigma^{\rm obs}_i}{\sigma^{\rm dressed}_i} =\frac {\int \sigma^{\rm dressed}_i(s(1-x))F(x,s)dx}{\sigma^{\rm dressed}_i(s)}},
\end{equation}
where $F(x,s)$ is the QED radiator function with an accuracy of 0.1\%~\cite{isr}, $x=2E_{\gamma}/\sqrt{s}$ the scaled ISR photon energy, and ${\sigma^{\rm obs}_i}={N^{\rm signal}_i}/( {{{\cal L}\cdot \epsilon_i \cdot\BR_i}})$ the observed cross section.

\begin{figure*}[htbp]
  \centering
  \subfigure[]{
  \label{Fig3.sub.1}
  \includegraphics[width=0.3\textwidth]{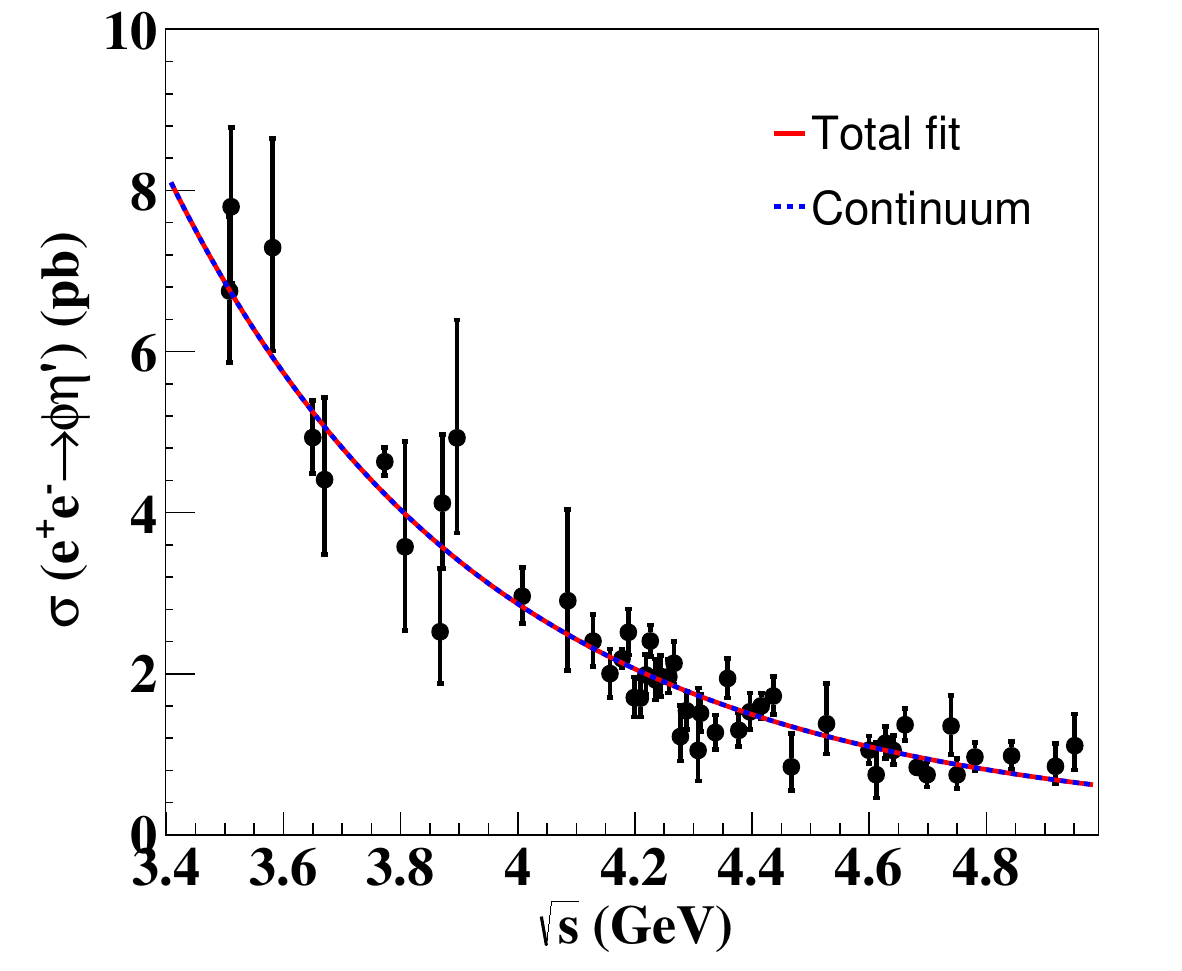}}
  \subfigure[]{
  \label{Fig3.sub.2}
  \includegraphics[width=0.3\textwidth]{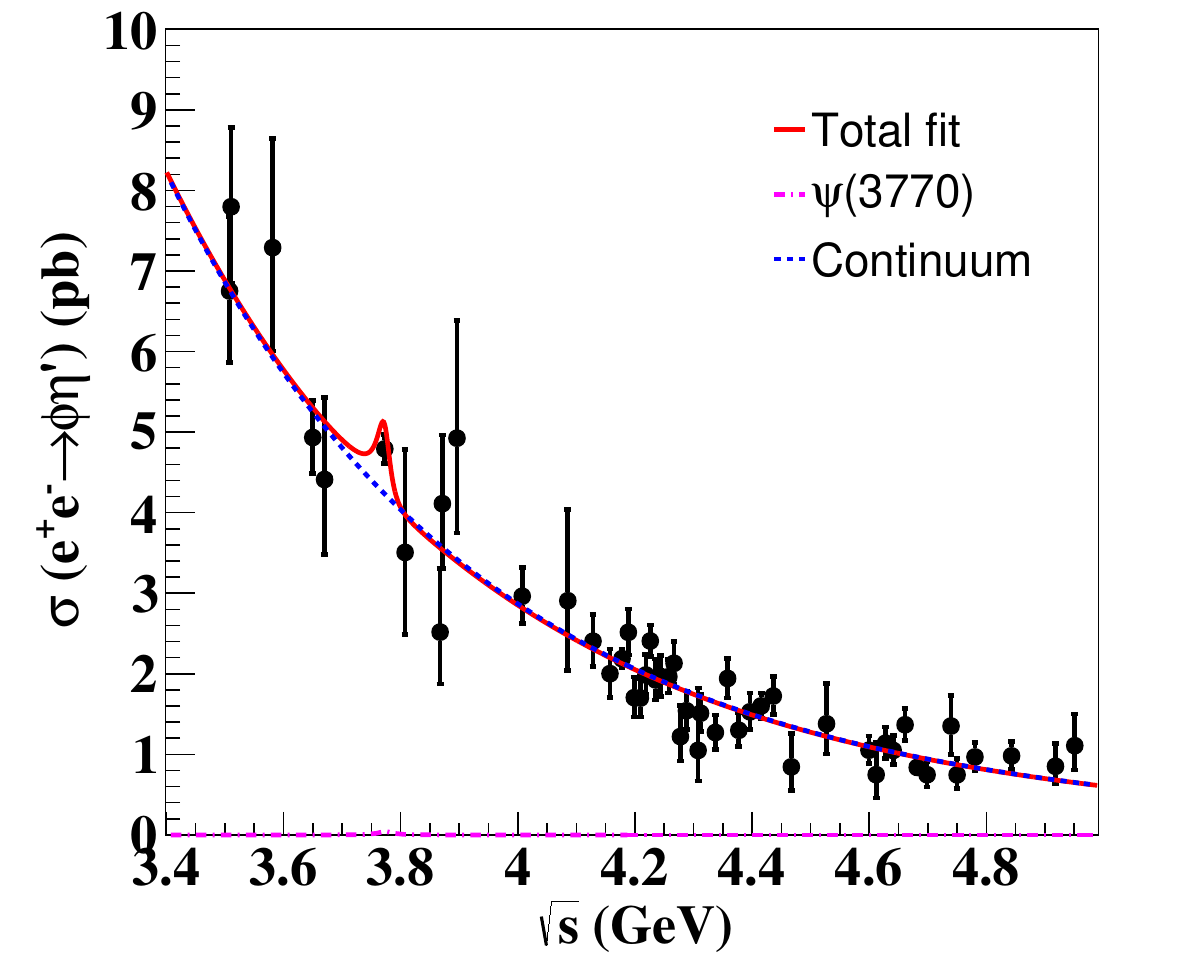}}
  \subfigure[]{
  \label{Fig3.sub.3}
  \includegraphics[width=0.3\textwidth]{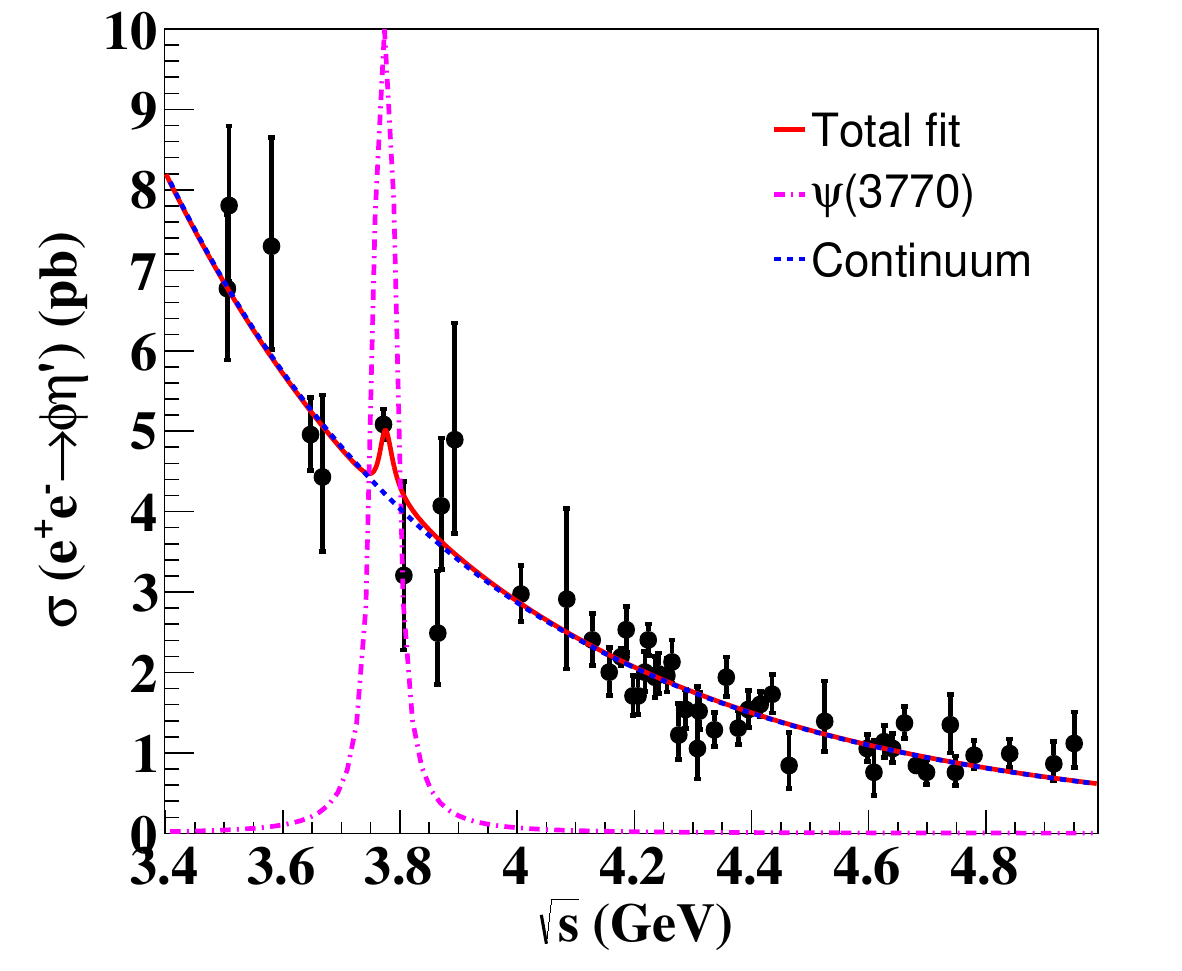}}
\caption{Dressed cross sections of the \eetophietap process and the fits under different assumptions: (a) continuum amplitude only; (b) coherent sum of continuum and $\pspp$ amplitudes, the solution with $\Phi=2.0$~rad, $\BR_{\phi\etap}=4.6\times 10^{-6}$, and the significance of the \psipptophietap~decay is $1.5\sigma$; (c) coherent sum of continuum and $\pspp$ amplitudes, the solution with $\Phi=4.7$~rad, $\BR_{\phi\etap}=1.9\times10^{-3}$, and the significance of the \psipptophietap~decay is $1.6\sigma$. The black dots with error bars are data. The red solid lines are the total fits, the magenta dot dashed lines represent the $\pspp$ component, and the blue dashed lines the continuum process. }\label{pic:fittingcs}
\end{figure*}

The cross sections we get from each of the modes are compatible with each other, and the combined dressed cross section of these two modes is calculated with
\begin{equation}\label{equ:445}
 {\sigma^{\rm dressed}_{\rm com} = \frac {N^{\rm signal}_{\rm com}} {{\cal L} \cdot (1+\delta)\cdot \sum\limits_i{\epsilon_i \cdot \BR_i}} },
\end{equation}
where $N^{\rm signal}_{\rm com}$ is the total number of signal events in the two modes. To get more accurate ISR factors, an iterative procedure is used~\cite{iteration}. The results of the cross sections are summarized in Table~\ref{tab:DressedCS-0.5pi} and shown in Fig.~\ref{pic:fittingcs}. We can see that the cross sections range between 0.7 and 7.5 pb and decrease gradually as $\sqrt s$ increases. There is no visible structure from the contribution of a vector charmonium or a charmonium-like state in this energy region.

Least-chi-square fits to the measured dressed cross sections are performed to describe the continuum line shape and to search for the \psipptophietap~decay. The $\chi^{2}$ is given by
\begin{equation}\label{equ:58}
 \chi^{2} =\sum_{j} {\frac{(\sigma_{j}-\sigma_{j}^{\rm fit})^{2}} {\delta_{j}^{2}} },
\end{equation}
where $\sigma_{j}$, $\sigma_{j}^{\rm fit}$, and $\delta_{j}$ are the measured dressed cross section, the fitted cross section, and the statistical uncertainty of the measured cross section of the $j$th energy point, respectively.

The cross sections are described with a coherent sum of the continuum and the $\pspp$ amplitudes
 \begin{equation}\label{equ:44}
 \sigma^{\rm dressed}(\sqrt s) = \left |\frac{a}{({\sqrt s})^{n}}\sqrt {{\rm PS}(\sqrt s)}
                                         + e^{i\Phi}{\rm BW}(\sqrt s)\right |^{2},
\end{equation}
where the constants $a$ and $n$ describe the magnitude and slope of the continuum process, $\Phi$ is the relative phase between the continuum and the $\pspp$ amplitudes, ${\rm PS}(\sqrt s)=q^3(\sqrt s)/s$ is the P-wave phase space factor~\cite{int1}, and $q(\sqrt s)$ is the momentum of $\phi$ and $\eta'$ in the $\EE$ c.m. frame.

The $\pspp$ resonance is parameterized with a Breit-Wigner (${\rm BW}$) function
 \begin{equation}\label{equ:45}
 {\rm BW(\sqrt s)}      =\frac{\sqrt {12\pi\Gamma_{\EE}\Gamma_{\rm tot}\BR_{\phi\etap}}}{s-M^{2}+iM\Gamma_{\rm tot}} \sqrt {\frac {{\rm PS}(\sqrt s)}{{\rm PS}(M)}},
\end{equation}
where $M$, $\Gamma_{\rm tot}$, and $\Gamma_{\EE}$ are the mass, width, and electronic partial width of $\pspp$, and all of them are set to their PDG values~\cite{int1}. $\BR_{\phi\etap}$ is the branching fraction of \psipptophietap.

For the data samples far from $\sqrt s=3.773$~GeV, the interference between the $\pspp$ and continuum amplitudes can be ignored.  Without the data sample at $\sqrt s=3.773$~GeV, the pure continuum process ($\sigma^{\rm dressed}(\sqrt s) = |\frac{a}{(\sqrt{s})^{n}}\sqrt {\rm PS(\sqrt s)}|^{2})$ is used to fit the dressed cross sections.
From this fit, $a=1.97 \pm 0.40$ (${\rm GeV}^{n-0.5}{\rm pb}^{0.5}$) and $n=4.35 \pm 0.14$ are obtained,
where the uncertainties are statistical only.
The systematic uncertainties of the cross section measurements (see below) have negligible effect on $n$.
The fit result is shown in Fig.~\ref{pic:fittingcs}(a).

\begin{figure}[ht]
  \centering
  \subfigure[]{
  \label{Fig4.sub.1}
  \includegraphics[width=0.45\textwidth]{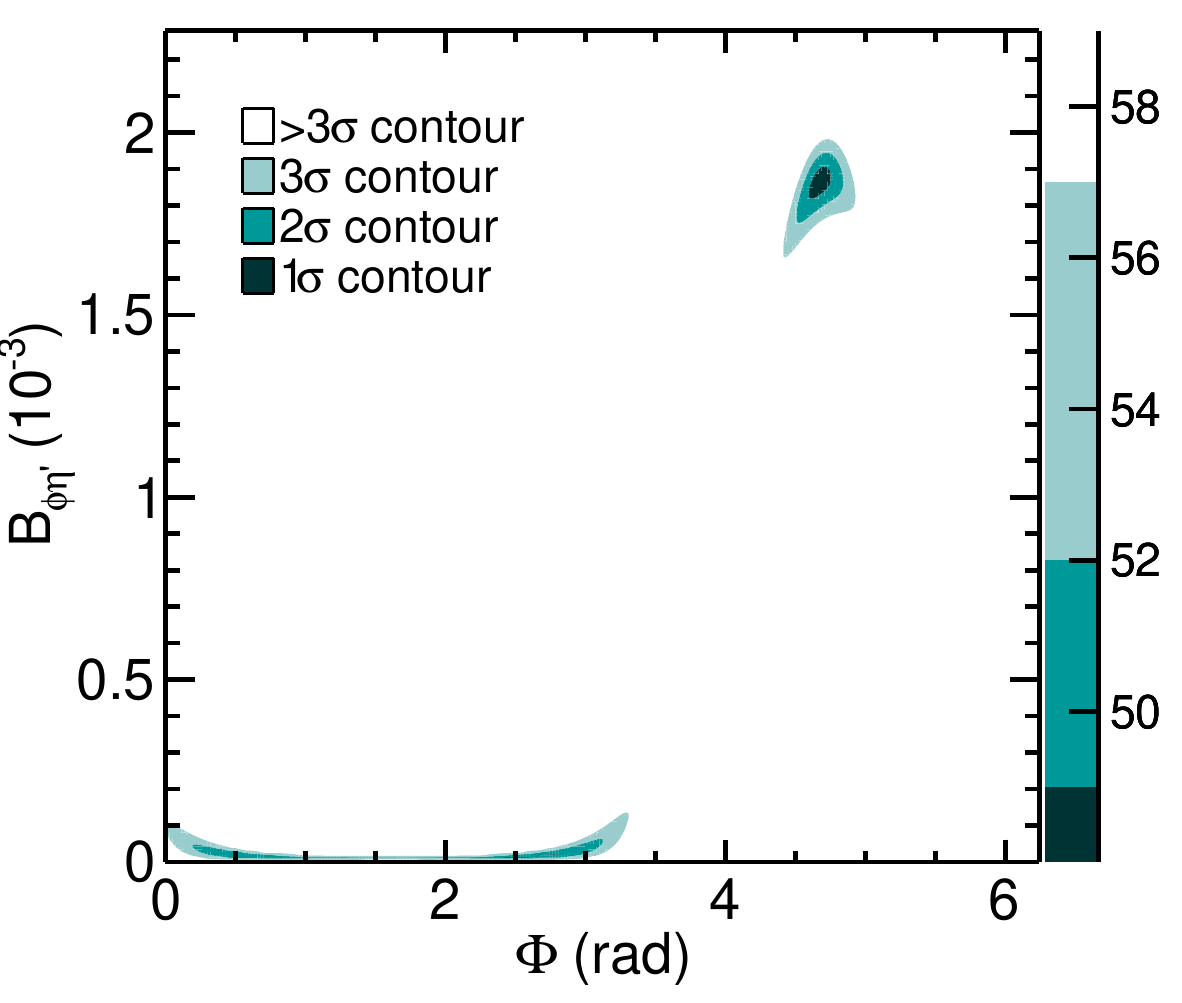}}
  \subfigure[]{
  \label{Fig4.sub.2}
    \includegraphics[width=0.45\textwidth]{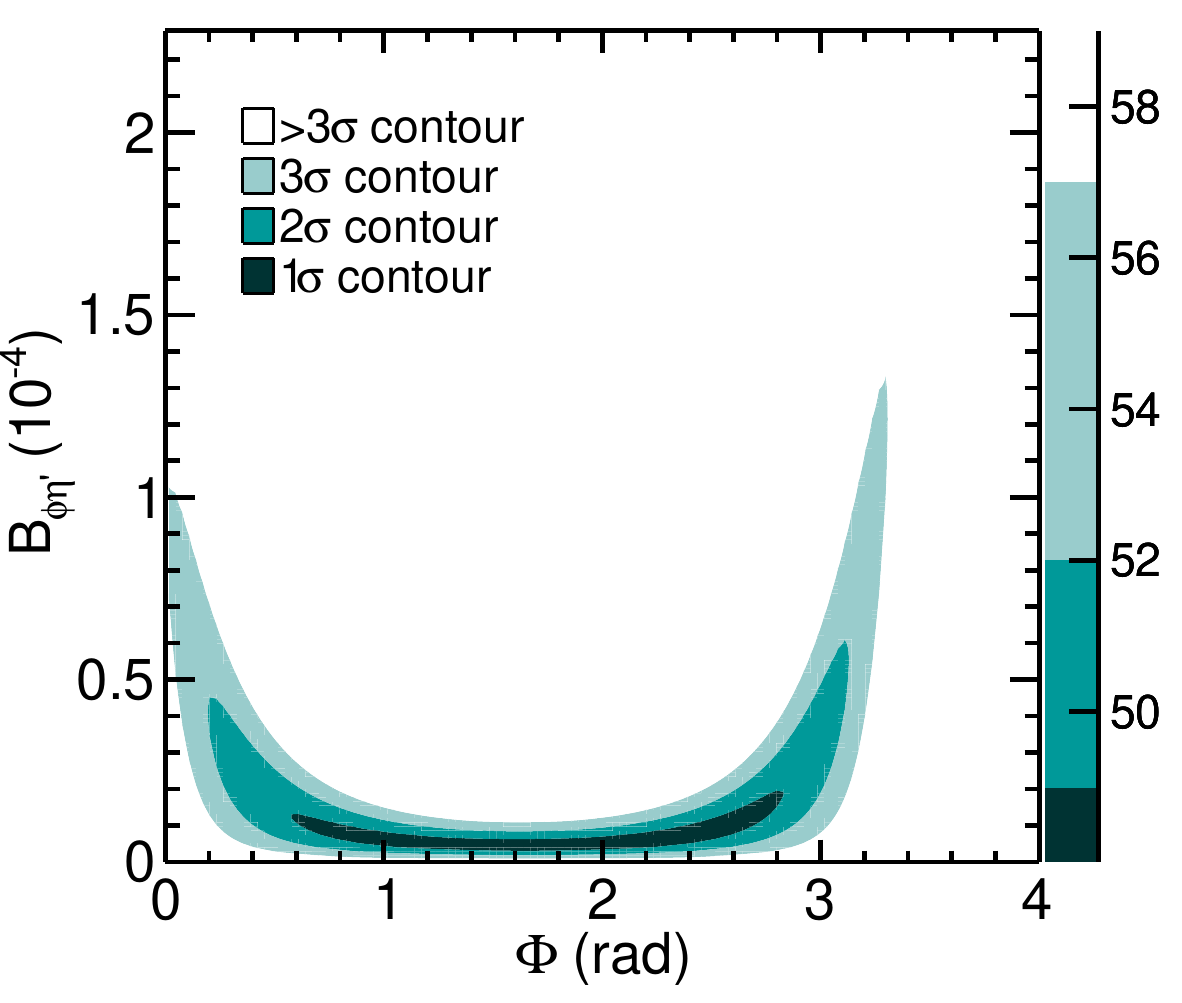}}
\caption{The two-dimensional scan result of the parameters $\BR_{\phi\etap}$ and $\Phi$. In (a), the $y$-axis is from 0 to $2.3\times10^{-3}$, and the upper right contour corresponds to Fig.~\ref{pic:fittingcs}(c) and the bottom
contour corresponds to Fig.~\ref{pic:fittingcs}(b), respectively; in (b), the $y$-axis is from 0 to $2.3\times10^{-4}$ (Plot (b) is the subset of plot (a)). The color in the figure represents the value of $\chi^2$ in different $\sigma$ intervals defined in Eq.~(\ref{equ:58}).}\label{pic:scanpar}
\end{figure}

Due to the lack of high integrated luminosity data in the vicinity of the $\pspp$, the fit including the $\sqrt{s}=3.773$~GeV data
and the resonance contribution is not stable. A two-dimensional parameter scan of  $\BR_{\phi\etap}$ and $\Phi$ is performed
under all possible line shape assumptions. In the scan, the parameters $a$ and $n$ describing the continuum amplitudes
are fixed to the central values obtained from the continuum-only fit. The resonance parameters of $\pspp$, $M$, $\Gamma_{\rm tot}$, and $\Gamma_{\EE}$ are fixed to the corresponding PDG values~\cite{int1}. We scan $\BR_{\phi\etap}$ and $\Phi$ from 0 to $2.3\times10^{-3}$ and from 0 to $2\pi$ in steps of $7.6\times 10^{-6}$ and 0.02, respectively. Figure~\ref{pic:scanpar} shows the scanning result, where the color scale represents the value of $\chi^{2}$ defined in Eq.~\eqref{equ:58}. The $i$th $\sigma$ contour is defined by $\chi^2=\chi^2_{\rm min} + i^2$~\cite{sta}, where $\chi^2_{\rm min}$ is the minimal $\chi^2$ value.
The probability that both parameters simultaneously take values within the one standard deviation ($1\sigma$) contour is 39.3\%. The fit result corresponding to the minimum in the scan (shown in Fig.~\ref{pic:scanpar}) is shown in Fig.~\ref{pic:fittingcs}(c).
The parameters $\BR_{\phi\etap}$ and $\Phi$ corresponding to the minimum $\chi^2$ value is ($1.9\times10^{-3}$,~4.7~rad), and the corresponding $\chi^2_{\rm min}$ is 48.0, which is slightly smaller than the $\chi^{2}$ of Fig.~\ref{pic:fittingcs}(b) ($\chi^{2}$ = 48.2). Therefore, these two solutions correspond to different line shapes and do not originate from the multiple-solution problems discussed in Ref.~\cite{Zhu:2011ha}.
The significance of the $\pspp\to \phi\etap$ decay is less than $2\sigma$ by comparing the fit $\chi^2$s ($\Delta~{\chi^2}=4.4$) with and without including the $\pspp$ amplitude and taking the change of the
number of degrees of freedom ($\Delta {\rm ndf}=2$) into account.
For the fitted parameter of $\BR_{\phi\etap}=1.9\times10^{-3}$, the cross section of \psipptophietap\ is of the same order of magnitude as that of the continuum process, as shown in Fig.~\ref{pic:fittingcs}(c). Due to destructive interference, a small structure appears above the expectation of the continuum process, which has also been observed in $e^+e^-\to\Lambda\bar{\Lambda},~\omega\pi^0$ and $\omega\eta$ processes~\cite{BESIII:2022tjc, BESIII:2021ccp, BESIII:2022zxr}.
Considering the fact that the cross section of $\pspp\to$ light~hadrons is much smaller than the continuum production of $\EE\to$ light~hadrons, we exclude the non-physical solution where the cross section is compatible to the continuum production. The physical solution, which lies within the $1\sigma$ contour range of $\BR_{\phi\etap}\in[0.3, 1.9]\times10^{-5}$ and $\Phi\in [0.6, 2.8]$ rad, is shown in Fig.~\ref{pic:scanpar}(b).

\section{SYSTEMATIC UNCERTAINTY}\label{sec:sys unc}
The systematic uncertainties of the cross section measurements are classified into two categories: correlated terms and uncorrelated terms. Correlated terms are common to the two $\etap$ decay modes, including the luminosity measurements, the differences between data and MC simulation for the tracking efficiency, photon reconstruction efficiency, the branching fraction of the $\phi$ decay, and the mass window for $\phi$ candidate. Uncorrelated terms are different in each $\etap$ decay mode, including the branching fractions of $\etap$ and $\eta$ decays, the $\eta$ reconstruction efficiency, the 4C/5C kinematic fits, the mass window for $\etap$ candidate, the signal shape and the background shape. These two categories of uncertainties are discussed in detail below.

The integrated luminosities of the data samples used in this study are measured using large angle Bhabha scattering events, with an uncertainty ranging within (0.5-1)\%~\cite{BESIII:2015qfd,BESIII:2022xyz,BESIII:2022xyz2}.  We conservatively take 1\% as the corresponding systematic uncertainty.
The uncertainty of the difference in the tracking efficiencies between data and MC simulations is estimated to be 1\% per track~\cite{track1, track2}. The uncertainty of the photon reconstruction efficiency is determined through dedicated studies~\cite{sys74,sys75} and is found to be 1\% per photon. The uncertainties related to the branching fractions of \phitokk, \etaptogammapipi, \etaptoetapipi, and \etatogammagamma\ are taken from PDG~\cite{int1}. The uncertainty caused by the $\eta$ reconstruction is determined by using a high purity control sample of $J/\psi \to\eta p \bar{p}$ decays~\cite{BESIII:2010tfr}. The study utilized the $\eta$ yields revealed in the $\gamma\gamma$ mass spectrum with or without $\eta$ selection requirements to determine the $\eta$ reconstruction efficiency.
Given the negligible statistical error of the test sample,
the observed 1\% difference of the $\eta$ reconstruction efficiencies between data and MC simulations is taken as the systematic uncertainty.

To address the uncertainty caused by the 4C/5C kinematic fit, we adjust the track helix parameters of charged tracks in the MC simulation so that it better describes the momentum spectra of the data~\cite{sys7}. In this study, we utilize the efficiency after the helix correction to obtain the nominal results. The difference in the MC efficiencies before and after the correction is taken as the systematic uncertainty.

The efficiencies obtained from MC simulation are corrected according to the data-MC difference in efficiencies obtained with the control sample of \psiptophietap~decays. The correction factor $f^{v}$ is defined as
 \begin{equation}\label{equ:1}
   f^{v}= \eff_{\rm MC}^{v}/\eff_{\rm data}^{v},
 \end{equation}
with
 \begin{equation}\label{equ:2}
   \eff^{v}_{\rm data(MC)}= N^{v}_{\rm data(MC)}/M^{v}_{\rm data(MC)},
 \end{equation}
where the subscripts ``MC" and ``data" represent MC and data samples, respectively,
$N^{v}_{\rm data(MC)}$ is the number of events in the signal region of a selection criterion $v$,
and $M^{v}_{\rm data(MC)}$ is the number of events in the full range of $v$.
The relative uncertainty of the correction factor $f^{v}$ is calculated
as a quadratic sum of the relative statistical errors of $\eff^{v}_{\rm data}$ and $\eff^{v}_{\rm MC}$.

A correction will not be applied if the correction factor $f^{v}$ differs from unity by less than its uncertainty;
otherwise the MC efficiency will be corrected as $\eff = \eff / f^{v}$, and $\sigma_{f^{v}}$ will be taken as the systematic uncertainty.
The systematic uncertainties due to the $\phi$ and $\etap$ mass windows are estimated using the control sample of \psiptophietap
(with \etaptogammapipi and \etaptoetapipi~(\etatogammagamma) for modes~I and~II, respectively).
The correction factors $f^{v}$ are $1.0072 \pm 0.0019$ for the $\phi$ mass window,
and $1.0065 \pm 0.0032$ (mode I) and $1.0104 \pm 0.0045$ (mode II) for the $\etap$ mass window.

For mode I, the parameters of the Gaussian functions convolved with the signal MC shape to describe the data are fixed at those obtained at $\sqrt s$ = 3.670, 3.867, 3.871, 4.278, 4.308, 4.467, 4.527, 4.699, 4.740, 4.918 and 4.951~GeV.
We vary the nominal mean and standard deviation of the Gaussian function by $\pm1\sigma$, and the maximum difference in the signal yields resulting from the various Gaussian function parameters is used to estimate the systematic uncertainty caused by the signal shape. We weight the systematic uncertainty of the signal shape by the luminosity of the data samples and obtain an estimated uncertainty of 1.6\%. For mode II, because the background is relatively small, the Gaussian function parameters in all data samples are free in the fit, and there is no systematic uncertainty caused by the signal shape.
The uncertainty associated with the background shape is estimated by changing from a first-order Chebyshev polynomial to a second-order one. To avoid any influence from statistical uncertainty, high statistical data samples at $\sqrt s$ = 3.773, 4.178, 4.420, and 4.680~GeV are used to evaluate the systematic uncertainty caused by the parameterization of the background shape. The largest differences in the signal yields between different parameterizations of the background shape are taken as the systematic uncertainties, which are 1.4\% and 2.7\% for modes~I and~II, respectively.

Due to correlations between the two modes, the combined $i$th item of the systematic uncertainty $\xi^{i}$ is calculated with
\begin{equation}\label{equ:554}
 \xi^{i}=\frac{\sqrt{(w_{\rm I}\xi_{\rm I}^{i})^{2}+(w_{\rm II}\xi_{\rm II}^{i})^{2}+
         2w_{\rm I}w_{\rm II}\rho_{\rm I,II}\xi_{\rm I}^{i}\xi_{\rm II}^{i} }}{w_{\rm I}+w_{\rm II}},
\end{equation}
where $w_{\rm I}=\BR$(\etaptogammapipi)$\cdot\epsilon_1$, and $w_{\rm II}=\BR$(\etaptoetapipi)$\cdot\BR$(\etatogammagamma)$\cdot\epsilon_2$. Here, $\xi_{\rm I}^{i}$ and $\xi_{\rm II}^{i}$ are the systematic uncertainties for modes~I and~II listed in Table~\ref{tab:TotalSysUncer}, respectively, and $\rho_{\rm I,II}$ is the correlation coefficient between modes~I and~II. For the systematic uncertainties caused by the luminosity measurements, the tracking efficiency, the photon reconstruction efficiency, the branching fraction of the $\phi$ decay, and the mass window for $\phi$ candidate, $\rho_{\rm I,II}$ is taken as 1, for the other sources as 0.

\begin{table}[!htbp]
\caption{The systematic uncertainties for the cross section measurements.}\label{tab:TotalSysUncer}\centering
\begin{tabular}{cccc}
\hline\hline
  \multirow{2}{*}{Source}   &\multicolumn{3}{c}{Systematic uncertainty (\%)}      \\
                          &\etaptogammapipi  &\etaptoetapipi  &Combination   \\
\hline
 Luminosity                   &{1.0}   &{1.0}    &1.0        \\
 Tracking                     &{4.0}   &{4.0}    &4.0        \\
 Photon                       &1.0     &2.0      &1.3     \\
 $\BR$(\phitokk)              &{1.0}   &{1.0}    &1.0    \\
 $\BR$(\etaptogammapipi)      &1.4     &-        &1.0   \\
 $\BR$(\etaptoetapipi)        &-       &1.2      &0.4   \\
 $\BR$(\etatogammagamma)      &-       &0.5      &0.2    \\
 $\eta$ reconstruction        &-       &1.0      &0.3         \\
 Kinematic fit                &1.2     &0.7      &0.9    \\
 Mass window of $\phi$        &0.2     &0.2      &0.2      \\
 Mass window of $\etap$       &0.3     &0.5      &0.3     \\
 Signal shape                 &1.6     &-        &1.1    \\
 Background shape             &1.4     &2.7      &1.3    \\
\hline
Total                         &5.2     &5.7      &5.0    \\
\hline\hline
\end{tabular}
\end{table}

Table~\ref{tab:TotalSysUncer} summarizes all the systematic uncertainties related to the cross section measurements for the individual decay modes and the combined one. The overall multiplicative systematic uncertainties are obtained by adding all systematic uncertainties in quadrature assuming they are independent.

\section{\boldmath UPPER LIMIT ON THE BRANCHING FRACTION of \psipptophietap}
There is no obvious structure in the dressed cross sections of $e^{+}e^{-}\to\phi\etap$, as shown in Fig.~\ref{pic:fittingcs}. Based on the Bayesian method~\cite{upp1}, the upper limit of the $\BR_{\phi\etap}$ at the 90\% confidence level (C.L.) is calculated with the systematic uncertainty taken into account.
The least-chi-square fits are performed on the dressed cross sections of~\eetophietap, and the fitting estimator $Q^{2}$ is constructed as

\begin{equation}\label{equ:599}
\small
\begin{aligned}
Q^{2} =\sum_{j} {\frac{(\sigma_{j}-\sigma_{j}^{\rm fit})^{2}} {\zeta_{j}^{2}}},
 \end{aligned}
\end{equation}
where $\sigma_{j}$, and $\sigma_{j}^{\rm fit}$ are defined according to Eq.~(\ref{equ:58}), and $\zeta_{j}$ includes the statistical uncertainty and uncorrelated systematic uncertainty of the measured cross sections, which are summed in quadrature assuming they are independent.

We vary the parameters $\BR_{\phi\etap}$ to get the distribution of $Q^{2}$. Then the likelihood distribution (${\rm L^{\prime}}$) as a function of the parameters $\BR_{\phi\etap}$ is constructed as ${{\rm L^{\prime}}(\BR_{\phi\etap}) = e^{-0.5Q^{2}}}$. The normalized likelihood distribution is smeared with a Gaussian function, with a mean $\eff_0$ and a standard deviation $\delta_{\eff} = \eff\times\delta^{cor}_{\rm sys}$, where $\delta_{\eff}$ and $\delta^{cor}_{\rm sys}$ are the absolute and relative correlated systematic uncertainties, respectively. The normalized likelihood is defined as
\begin{equation}\label{equ:59}
\small
\begin{aligned}
 &{{\rm L}(\BR_{\phi\etap}) = \int_{0}^{1} {\rm L^{\prime}}({\frac{\eff}{\eff_{0}}\cdot\BR_{\phi\etap}})\cdot\frac{1}{\sqrt{2\pi}\delta_{\eff}}e^{-\frac{{(\eff-\eff_{0})}^{2}}{2\delta_{\eff}^{2}}}}{\rm d\eff},
 \end{aligned}
\end{equation}
where $\eff_{0}$ is the efficiency obtained from the MC simulation.

The parameters $\BR_{\phi\etap}$ at the 90\% C.L. ($\BR_{\phi\etap}^{\rm up}$) are determined as
 \begin{equation}\label{equ:10}
 \begin{aligned}
  \int_{0}^{\BR_{\phi\etap}^{\rm up}}{\rm L}(\BR_{\phi\etap}){\rm d\BR_{\phi\etap}} =
   0.9\int_{0}^{\infty}{\rm L}(\BR_{\phi\etap}){\rm d\BR_{\phi\etap}}.
   \end{aligned}
 \end{equation}

\begin{figure}[!ht]
  \centering
  \includegraphics[width=0.45\textwidth]{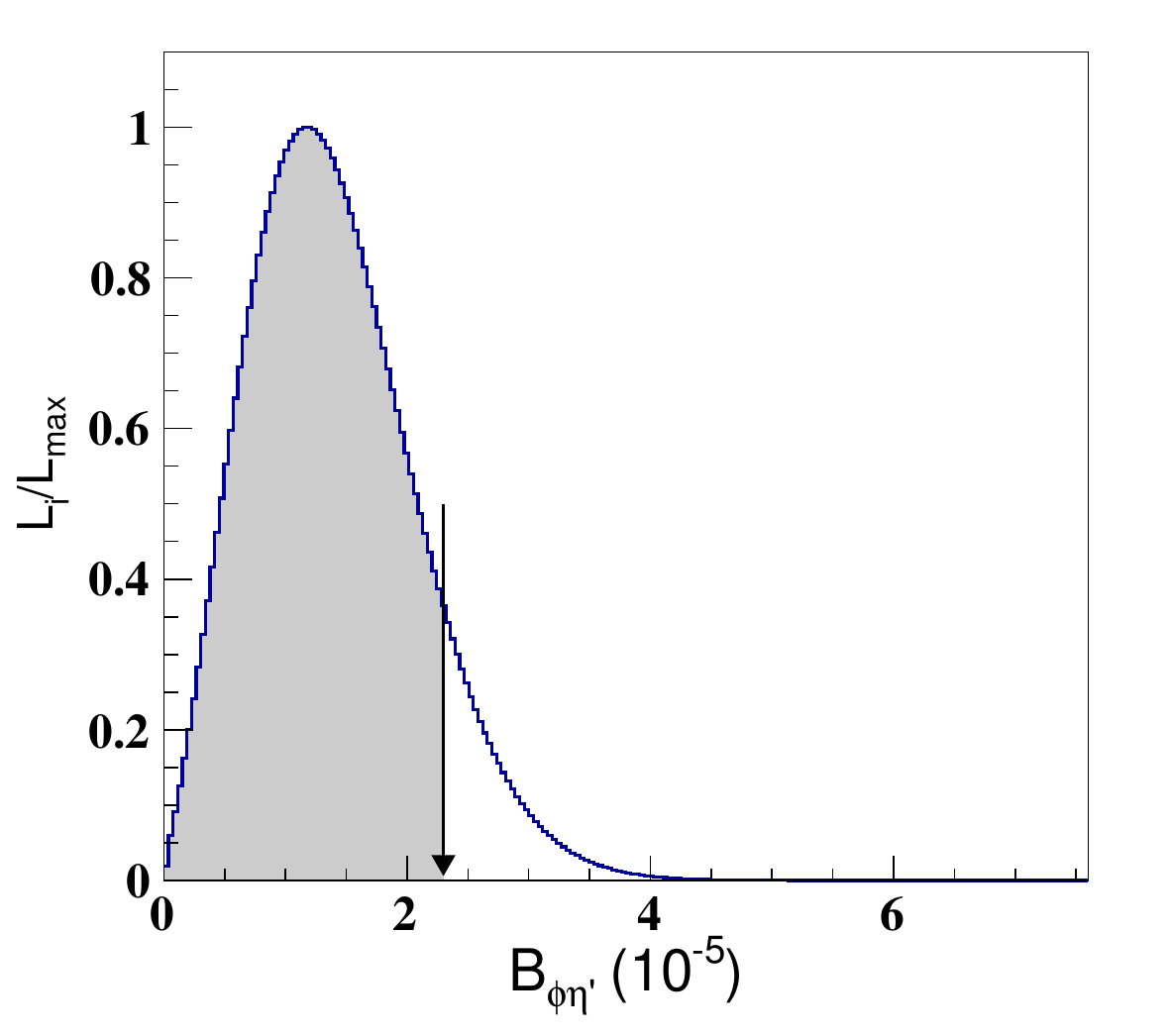}
\caption{The distribution of the likelihood versus $\BR_{\phi\etap}$, with the phase $\Phi$ fixed to 2.81~rad. The black arrow shows the result corresponding to the 90\% C.L.}\label{pic:scan-phi}
\end{figure}

Considering the effect on the upper limit from the uncertainties of the mass and width measurements, we vary the nominal mass and width by $\pm 1\sigma$ and choose the largest upper limit of $\BR_{\phi\etap}$ among these combinations.
The maximum of $\BR_{\phi\etap}$ is near $\Phi=2.81$~rad within the $1\sigma$ contour shown in Fig.~\ref{pic:scanpar}.
In order to obtain more conservative upper limits, the phase $\Phi$ is fixed to 2.81~rad when determining the upper limit of $\BR_{\phi\etap}$. The normalized likelihood distribution is displayed in Fig.~\ref{pic:scan-phi}, and an upper limit of $2.3\times 10^{-5}$ is obtained for $\BR_{\phi\etap}$ for the charmless decay of $\pspp\to\phi\eta'$.

\section{SUMMARY and discussions}
The measurement of the \eetophietap~cross sections at center-of-mass energies from 3.508 to 4.951 GeV have been reported.
The results in Table~\ref{tab:DressedCS-0.5pi} show that the cross sections range from 0.71$\pm0.20$ to 7.46$\pm0.95$ pb, which is not consistent with $\sim60$ pb predicted by the NJL model~\cite{Bystritskiy:2008pr}. The power parameter, $n=4.35\pm 0.14$, is obtained from the fit to the cross sections. This value is in good agreement with the prediction of the NJL model ($n=3.5\pm 0.9$)~\cite{Bystritskiy:2008pr}, where the uncertainty is solely systematic. However, it should be noted that the accuracy of the NJL model is (20-30)\% for the calculation of the cross sections of the $\EE\to$ VP process.
Our present study can therefore offer valuable insights to advance the development of the NJL model and enhance its precision in calculating the cross sections of the VP process.

By fitting the measured cross sections, we have also searched for the charmless decay $\pspp\to\phi\etap$ and found that its statistical significance is less than $2\sigma$. The upper limit for the parameters $\BR_{\phi\etap}$ of the $\pspp$ line shape has been determined to be $2.3\times 10^{-5}$, which is in agreement with the model prediction of $(0.5-3.5)\times10^{-5}$~\cite{int45, int46}.
This suggests that the S- and D-wave charmonium states mixing scheme may play a role in explaining the ``$\rho\pi$ puzzle'' in $J/\psi$, $\psi(2S)$, and $\psi(3770)$ charmless decays. This study, together with previous searches for exclusive $\pspp$ charmless decays, provides valuable insights into the nature of the $\pspp$, but the large $\pspp$ ${\rm non}$-$D\bar{D}$ decay rate remains a puzzle. In the future, a fine energy scan
of the $\pspp$ resonance would be necessary to determine the presence and magnitude of its charmless decays.

\section{Acknowledgment}

The BESIII Collaboration thanks the staff of BEPCII and the IHEP computing center for their strong support. This work is supported in part by National Key R\&D Program of China under Contracts Nos. 2020YFA0406300, 2020YFA0406400; National Natural Science Foundation of China (NSFC) under Contracts Nos. 11635010, 11735014, 11835012, 11935015, 11935016, 11935018, 11961141012, 12022510, 12025502, 12035009, 12035013, 12061131003, 12192260, 12192261, 12192262, 12192263, 12192264, 12192265, 12221005, 12225509, 12235017; the Chinese Academy of Sciences (CAS) Large-Scale Scientific Facility Program; the CAS Center for Excellence in Particle Physics (CCEPP); Joint Large-Scale Scientific Facility Funds of the NSFC and CAS under Contract No. U1832207; CAS Key Research Program of Frontier Sciences under Contracts Nos. QYZDJ-SSW-SLH003, QYZDJ-SSW-SLH040; 100 Talents Program of CAS; The Institute of Nuclear and Particle Physics (INPAC) and Shanghai Key Laboratory for Particle Physics and Cosmology; ERC under Contract No. 758462; European Union's Horizon 2020 research and innovation programme under Marie Sklodowska-Curie grant agreement under Contract No. 894790; German Research Foundation DFG under Contracts Nos. 455635585, Collaborative Research Center CRC 1044, FOR5327, GRK 2149; Istituto Nazionale di Fisica Nucleare, Italy; Ministry of Development of Turkey under Contract No. DPT2006K-120470; National Research Foundation of Korea under Contract No. NRF-2022R1A2C1092335; National Science and Technology fund of Mongolia; National Science Research and Innovation Fund (NSRF) via the Program Management Unit for Human Resources \& Institutional Development, Research and Innovation of Thailand under Contract No. B16F640076; Polish National Science Centre under Contract No. 2019/35/O/ST2/02907; The Swedish Research Council; U. S. Department of Energy under Contract No. DE-FG02-05ER41374.

%

\nocite{*}

\end{document}